\documentclass[aoas,MSNbibl,nameyear,seceqn,dvips]{arximspdf}
\usepackage{dcolumn}
\usepackage{graphicx}

\doi{10.1214/11-AOAS477}
\volume{5}
\issue{3}
\pubyear{2011}
\firstpage{1876}
\lastpage{1892}

\makeatletter
\newcommand\independent{\protect\mathpalette{\protect\independenT}{\perp}}
\def\independenT#1#2{\mathrel{\rlap{$#1#2$}\mkern4.1mu{#1#2}}}
\newcolumntype{d}[1]{D{.}{.}{#1}}
\makeatother

\begin{document}
\begin{frontmatter}

\title{The potential for bias in principal causal effect estimation
when treatment received depends on~a~key covariate\thanksref{TITL1}} %
\runtitle{Principal effects with compliance-predictive covariate}
\thankstext{TITL1}{Supported from NIDA Grants 1-R01-DA025680 and 1-R01-DA016850
as well as NIMH Grants 5-P30-MH58017 and 1-P30-MH082760.}

\begin{aug}
\author[A]{\fnms{Corwin M.} \snm{Zigler}\corref{}\ead[label=e1]{czigler@hsph.harvard.edu}}
and
\author[B]{\fnms{Thomas R.} \snm{Belin}\ead[label=e2]{tbelin@mednet.ucla.edu}}
\runauthor{C. M. Zigler and T. R. Belin}
\affiliation{Harvard University and University of California, Los Angeles}
\address[A]{Harvard School of Public Health\\
Building 2, 4th Floor \\
655 Huntington Ave \\
Boston, Massachusetts 02115 \\
USA\\
\printead{e1}}

\address[B]{Department of Biostatistics\\
UCLA School of Public Health\\
51-267 Center for Health Sciences \\
Los Angeles, California 90095-1772\\
USA\\
\printead{e2}}
\end{aug}

\received{\smonth{11} \syear{2010}}
\revised{\smonth{4} \syear{2011}}

%
\begin{abstract}
Motivated by a potential-outcomes perspective, the idea of principal
stratification has been widely recognized for its relevance in settings
susceptible to posttreatment selection bias such as randomized clinical
trials where treatment received can differ from treatment assigned. In
one such setting, we address subtleties involved in inference for
causal effects when using a key covariate to predict membership in
latent principal strata. We show that when treatment received can
differ from treatment assigned in both study arms, incorporating a
stratum-predictive covariate can make estimates of the ``complier
average causal effect'' (CACE) derive from observations in the two
treatment arms with different covariate distributions. Adopting a
Bayesian perspective and using Markov chain Monte Carlo for
computation, we develop posterior checks that characterize the extent
to which incorporating the pretreatment covariate endangers estimation
of the CACE. We apply the method to analyze a clinical trial comparing
two treatments for jaw fractures in which the study protocol allowed
surgeons to overrule both possible randomized treatment assignments
based on their clinical judgment and the data contained a key covariate
(injury severity) predictive of treatment received.
\end{abstract}

%

\end{frontmatter}

\section{Introduction}
All-or-none treatment noncompliance in the context of a randomized two-arm
clinical trial is perhaps the simplest and most common example in
health-sciences research of potential confounding by a
posttreatment variable. One strategy to address confounding of
treatment receipt with individual characteristics is the use of an
instrumental-variable method [\citet{mcclellandoes1994}]
which has been linked to a potential-outcomes perspective on
causal inference [\citet{angristidentification1996}; \citet{imbensbayesian1997}; \citet{frangakisaddressing1999}].
More recently, strategies for addressing potential
confounding by posttreatment variables have been formalized using the
framework of principal stratification [\citet{frangakisprincipal2002}], a central challenge of which is the classification of
patients into latent subclasses, called principal strata, that
facilitate causal treatment comparisons.

In the context of treatment noncompliance, the target for inference is often the ``complier average
causal effect (CACE)'' [\citet{imbensbayesian1997}],
which compares treatment outcomes with control outcomes in
the principal stratum of ``compliers'' who would potentially receive
whichever treatment is randomly assigned (as distinct from other
principal strata where patients may always receive a particular
treatment). Such comparisons within principal strata are known as
``principal effects'' and permit causal interpretation. Knowledge of
membership in the stratum of compliers or in any other principal
stratum requires knowledge of patients' potential treatment receipts
under both possible randomized assignments, but this information will
never be observed in total for any individual in the population since
treatment received is only observed for the actually assigned treatment.

A battery of now-standard assumptions underlie methods for identifying
and estimating the CACE in settings framed as treatment noncompliance
[\citet{angristidentification1996}], but more recent attention
[e.g., \citet{hiranoassessing2000}; \citet{jouse2009}] has
been paid to the use of pretreatment covariates to increase precision
or relax exclusion restrictions. One line of research focuses on
settings where patients randomized to the control arm do not have
access to the active treatment, that is, settings where the entire
population would receive control if so assigned. The key feature of
these settings is that they allow patients who are assigned and receive
active treatment to be identified as compliers, which further allows
pretreatment covariates associated with membership in this stratum to
be used in identifying which patients randomized to control are
exchangeable with compliers. Specifically, these settings motivate
so-called ``two-stage'' approaches that first use pretreatment
covariates to estimate propensity scores
[\citet{rosenbaumcentral1983}] of membership in the complier
stratum, then estimate outcomes conditional on these so-called
``principal scores'' [\citet{follmanneffect2000};
\citet{hillsustained2003}; \citet{joffecompliance2003};
\citet{joffedefining2007}; \citet{jouse2009}]. Although some
previous research has framed the one-sided access to treatment as a
nonessential detail that merely simplifies exposition, we aim to
illuminate that added complexity can arise in more general settings
where noncompliance exists in both treatment arms.

When treatment received can deviate from treatment assigned in both
study arms, the use of pretreatment covariates to aid estimation of the
CACE is more complicated because no patient is known to belong to the
stratum of compliers, precluding estimation of a model such as a
propensity score model for membership in the complier stratum.
Joint-estimation methods that simultaneously model stratum membership
and outcomes have been employed in these settings
[\citet{hiranoassessing2000}; \citet{frangakisclustered2002}; \citet{barnardprincipal2003};
\citet{griffinapplication2008}; \citet{royprincipal2008}; \citet{gallopmediation2009}],
which typically consist of two underlying strata in addition to the
compliers: ``never-takers'' who would never receive the active
treatment, and ``always-takers'' who would always receive the active
treatment. Through use of standard assumptions that will be elaborated
later, stratum membership for patients who receive a treatment
different from that assigned can be regarded as having been revealed,
with such individuals being either never-takers or always-takers, and
covariates associated with membership in these two ``noncomplier''
strata can be identified. However, membership in the complier stratum
is never directly observed because patients who receive the assigned
treatment (and thus \textit{might} be compliers) generally represent
mixtures of compliers and never-takers (in the control arm) or
compliers and always-takers (in the treatment arm). Since pretreatment
covariates can only provide direct information about characteristics of
noncompliers, the role of such covariates in estimating the CACE is to
model which patients in the complier/noncomplier mixtures are
noncompliers, thus indirectly estimating the remaining portion of the
mixture to belong to the stratum of compliers.

In this article we employ a
joint-estimation method using a Gibbs sampling computational approach
[\citet{gemanstochastic1984}; \citet{gelfandsampling-based1990}]
in a setting where noncompliance exists in both
randomization arms. We aim to improve the estimate of the CACE through
incorporation of a compliance-predictive model that uses a key
covariate to select compliers from the complier/noncomplier mixtures.
Our novel contribution is a~detailed exposition of scenarios in which
observed data predict membership in the noncomplier strata in a way
that can select compliers in each treatment group from different
portions of the covariate distribution, potentially implying that the
estimated CACE is biased for the causal effect of treatment. After
introducing the motivating oral-surgery application in Section \ref
{surgerytrial}, Section \ref{definePO} formally defines a
potential-outcomes inference framework and the assumptions necessary
for estimation of the CACE. Section~\ref{model} develops the
compliance-predictive model and corresponding estimation procedure.
Section \ref{sims} uses simulated examples to illustrate some posterior
checks and illuminate the potential for bias resulting from the
compliance-predictive model, and Section \ref{dentdat} illustrates the
impact of using the key covariate to predict compliance status in the
oral-surgery setting. We conclude with a~discussion.

\section{Motivating oral-surgery clinical trial}\label{surgerytrial}
Our motivating example consists of 142 patients who were randomly
assigned to receive treatment for jaw fractures in
the form of Maxillomandibular Fixation (MMF, control) or Rigid Internal
Fixation (RIF, active treatment). The study aimed to investigate the
putative advantages of the increasingly-popular RIF over the more
traditional MMF in a patient population thought to be prone to
postoperative complications. A degree of clinical flexibility was
deemed essential to the protocol, allowing treatment decisions to
depart from the randomized treatment assignment if deemed necessary by
the treating surgeon. This clinical latitude gives rise to possible concerns
that more severely injured patients were disproportionately selected
into the more
aggressive treatment arm, as it is well accepted in the surgical
community that the MMF procedure, which is less expensive, is
appropriate for less severe injuries while the RIF procedure, which is
more resource-intensive, is appropriate for more severe injuries.
Although the exact rationale for treatment
decisions was not recorded, a~continuously-scaled measure of injury
severity ($\mathit{SEV}$) was calculated for each patient. This severity
measure, originally developed as the Mandible Injury Severity
Score (MISS) [\citet{shettymandible2007}], ranges from 0 (less severe) to 25
(extremely severe), and derives from anatomic and clinical
characteristics of the constituent jaw fractures. The outcome of
interest was a continuously-scaled General
Oral Health Assessment Index ($\mathit{GOHAI}$) [\citet{atchisongeneral1997}] measured at six months post-treatment, with higher values
suggesting better oral-health quality of life. In the face of
``noncompliance'' (i.e., surgical judgment overriding the treatment
assigned through the randomization protocol), one could conduct
intention-to-treat and as-treated analyses [\citet{shettybenefits2008}],
but the former addresses a~question that is arguably not
the only scientific question of interest, the latter can give rise to
bias in estimates of the treatment effect, and neither accounts for the
plausible effect that subjective treatment decisions had on the
analysis.\looseness=1

\section{Potential outcomes, principal strata and causal estimand}\label{definePO}
Definition of principal strata and causal estimands requires
development of a potential-outcomes framework, often called the
Rubin Causal Model [\citet{rubinbayesian1978}; \citet{hollandstatistics1986}]. Following previous
development in the setting of all-or-none
treatment noncompliance in a two-arm clinical trial
[\citet{angristidentification1996}],
we define potential outcomes and delineate the principal strata
that arise in our motivating setting. We then outline the
assumptions necessary for identifiability of the causal estimand of
interest, the CACE.\looseness=1

\subsection{Potential outcomes and principal strata}
First, we define the relevant potential outcomes inherent in this
clinical trial. Define $\mathbf{Z}$ as the vector of random
treatment assignments for all patients in the study, with
$i$th element $Z_i$ equal to $0$ for assignment to MMF and
$1$ for assignment to RIF. Let $\mathbf{D(Z)}$ be a vector with $i$th
element $D_i(\mathbf{Z})$ denoting the $i$th patient's
received treatment under assignment~$\mathbf{Z}$. Patients with
$D_i(\mathbf{Z})=0$ would receive treatment with MMF under
assignment $\mathbf{Z}$, while patients with $D_i(\mathbf{Z})=1$
would receive treatment with RIF under assignment~$\mathbf{Z}$.
Furthermore, we use $Y_i(\mathbf{Z},\mathbf{D})$ to denote a patient's
potential $ \mathit{GOHAI}$  with respect to
$\mathbf{Z}$ and~$\mathbf{D}$. We
adopt the stable unit treatment value assumption (SUTVA)
[\citet{rubinbayesian1978}] here to indicate no interference between patients, allowing
us to write
$D_i(\mathbf{Z})=D_i(Z_i)$ and $Y_i(\mathbf{Z},
\mathbf{D})=Y_i(Z_i)$.

Principal strata in this setting are defined by all four possible
values of the pair $(D_i(0), D_i(1))$. We call the principal
stratum of patients with $(\mbox{$D_i(0)=0$},D_i(1)=1)$ ``compliers'' who will
receive the assigned treatment regardless of which treatment is
assigned, and denote these patients as having $S_i=c$. Similarly, we
can call the
stratum with $(D_i(0)=0,D_i(1)=0)$ ``never-takers'' who will never
receive treatment with RIF, denoting these patients with $S_i=n$, and
the stratum of patients with
$(D_i(0)=1,D_i(1)=1)$ ``always-takers'' who will always receive
treatment with RIF, which we label with $S_i=a$. Finally, we define the
principal stratum of ``defiers''
as those with $(D_i(0)=1,D_i(1)=0)$, or those who will always
receive the treatment opposite of that assigned, with $S_i=d$.

Naturally, we observe only one component of $(D_i(0), D_i(1))$ and only
one component of $(Y_i(0), Y_i(1))$. To draw out this distinction
between observed and missing
components, we write $(D_i^{\mathit{obs}}, D_i^{\mathit{mis}})$
and $(Y_i^{\mathit{obs}}, Y_i^{\mathit{mis}})$, where the
superscripts \textit{obs}
and \textit{mis} denote the observed and missing potential
outcomes, respectively.

\vspace*{6pt}
\subsection{Assumptions for identifiability of causal estimands}
As no complete pair of potential outcomes is observable, we
require additional assumptions for identifiability of causal
estimands. In addition to SUTVA, we adopt a~monotonicity assumption
[\citet{imbensidentification1994}]
disallowing the existence of the principal
stratum of defiers, that is, there are no patients who would
receive MMF if assigned RIF but receive RIF if assigned MMF. This
setting with noncompliance
resulting from clinicians' judgment is unlikely to produce a
violation of the monotonicity assumption. The usefulness of
monotonicity lies in its implication that patients with
$D_i^{\mathit{obs}}=D_i(0)=1$ must belong to the stratum of always-takers
and those with $D_i^{\mathit{obs}}=D_i(1)=0$ must belong to the stratum of never-takers.
Stratum membership for those who received the assigned treatment
remains unidentified, as patients with $D_i^{\mathit{obs}}=D_i(0)=0$ represent a
mixture of compliers and never-takers, while those with
$D_i^{\mathit{obs}}=D_i(1)=1$ represent a mixture of compliers and
always-takers. The first three columns of Table~\ref{compsum} provide a
summary of the possible principal strata for patients with each
possible observed
pattern of $Z_i$ and~$D_i^{\mathit{obs}}$.

\begin{table}[t]
\tabcolsep=0pt
\caption{Possible principal strata for observed treatment assignment
and receipt patterns and summary statistics for $\mathit{SEV}$ and $\mathit{GOHAI}$ in
the motivating oral-surgery setting} \label{compsum}
\begin{tabular*}{\tablewidth}{@{\extracolsep{\fill}}lccccc@{}}
\hline
\textbf{Treatment} & \textbf{Treatment} & \textbf{Possible principal strata} &
 & \textbf{Mean (SD)} & \textbf{Mean (SD)} \\
\textbf{assigned,} $\bolds{Z_i}$ & \textbf{received,} $\bolds{D_i^{\mathit{obs}}(Z_i)}$ &
$\bolds{(D_i(0), D_i(1))}$ &$\bolds{n}$& $\bolds{\mathit{SEV}}$ & $\bolds{\mathit{GOHAI}}$ \\
\hline
0 & 0 & compliers or never-takers & 53 & 12.8 (2.7)
& 42.8 (12.1) \\
& & $(D_i(0)=0, D_i(1)=0$ or $1)$ \\[3pt]
0 & 1 & always-takers & \hphantom{0}9 & 14.0 (2.0) & 42.8
(11.9) \\
& & $(D_i(0)=1, D_i(1)=1)$ \\[3pt]
1 & 1 & compliers or always-takers & 40 & 13.2
(2.3) & 44.5 (12.1)\\
& & $(D_i(0)=1$ or $0, D_i(1)=1)$ \\[3pt]
1 & 0 & never-takers & 40 & 12.2 (3.0) & 41.7 (9.3)\\
& & $(D_i(0)=0, D_i(1)=0)$\\
\hline
\end{tabular*}
\vspace*{15pt}
\end{table}

Inference for causal effects in clinical trials with treatment
noncompliance typically relies on another assumption, known as the
exclusion restriction
[\citet{angristidentification1996}], stating that any effect
of treatment assignment, $Z$,
on the outcome, $Y$, must be via an effect of treatment~recei\-ved,~$D$.
After accounting for received treatment, random assignment no
longer affects $\mathit{GOHAI}$, or $(Y(z)|Z=z,D(z)=d) = (Y(z)|D(z)=d)$ for
$z=0,1$ and $d=0,1$.

With the above development, we define the CACE as the expected
difference in potential $\mathit{GOHAI}$ outcomes within the stratum of compliers:
\[
\mathit{CACE} = E[Y_i(1)-Y_i(0)|S_i=c].
\]

\vspace*{6pt}
\section{Bayesian models for the CACE with the compliance-predictive feature}\label{model}
We formulate our inference strategy with a phenomenological Bayesian
model following
\citet{imbensbayesian1997}. The model is phenomenological in the sense
described by \citeauthor{rubinbayesian1978} (\citeyear{rubinbayesian1978}, \citeyear{rubinmultiple1978}), where the
inference builds on potentially observable quantities even though not
all of the quantities will be observed. The relevant random variables
for each patient are $Z_i,
D_i(0), D_i(1),\allowbreak Y_i(0), Y_i(1)$ and
$X_i$, where $X_i$ denotes the $i$th patient's $\mathit{SEV}$. We consider
these random variables realizations from a joint
distribution, with $X_i$, $Z_i$, $D_i^{\mathit{obs}}$  and $Y_i^{\mathit{obs}}$
observed for each patient. Our goal is to model the conditional
distributions of $Y_i(z)$ conditional on principal stratum, which
requires integration over missing values as a result of the
unidentifiable mixtures over the latent $S_i$. This motivates a
Gibbs-sampling strategy that first samples the missing $S_i$, thereby
allowing assessment of the distributions of $Y_i(z)$ conditional on the
``complete compliance data'' consisting of subpopulations without
mixture components.

\subsection{Structure of Bayesian inference}\label{structure}
The joint distribution of the data can be factored as follows:
%
\begin{eqnarray}\label{joint}
f\bigl(\mathbf{Z, (Y(0),Y(1)), (D(0), D(1)), X}\bigr)& =& f(\mathbf{Z,Y,S,X})\nonumber\\
& =&
f(\mathbf{Y,S,X|Z})f(\mathbf{Z})\\
&=& f(\mathbf{Y,S,X})f(\mathbf{Z}),\nonumber
\end{eqnarray}
where the last equality holds due to randomization in the study design.
We facilitate Bayesian inference by writing the joint distribution of
$\mathbf{Y, S}$ and~$\mathbf{X}$ as the product of independently
identically distributed random variables conditional on a generic
parameter $\theta$ [\citet{definettitheory1974}], where we denote the prior distribution of~$\theta$ as
$p(\theta)$ and the posterior distribution of $\theta$ as
%
\begin{eqnarray}\label{fullposterior}
\quad&&p(\theta|\mathbf{Y^{\mathit{obs}}, D^{\mathit{obs}}, X, Z}) \nonumber\\[-8pt]\\[-8pt]
\quad&&\qquad \propto
p(\theta)\int\!\!\!\int\prod f(Y_i^{\mathit{obs}}, Y_i^{\mathit{mis}}, D_i^{\mathit{obs}},
D_i^{\mathit{mis}}, X_i|\theta)\,dY_i^{\mathit{mis}}\,dD_i^{\mathit{mis}}.\nonumber
\end{eqnarray}
As pointed out in \citet{frangakisclustered2002} and \citet{jinprincipal2008},
required integration over $\mathbf{D^{\mathit{mis}}}$ proves
computationally difficult in general, but as a result of randomization
$\mathbf{Y^{\mathit{mis}}}$ can be handled with standard randomization-based
tools. Furthermore, the difficult integration over $\mathbf{D^{\mathit{mis}}}$
leads us to consider the joint posterior of $(\mathbf{\theta,D^{\mathit{mis}}})$,
%
\begin{equation}\label{thetaSposterior}
\qquad p(\mathbf{\theta, D^{\mathit{mis}}|D^{\mathit{obs}},Y^{\mathit{obs}}, X, Z}) \propto
p(\theta)\prod f(D_i^{\mathit{obs}}, D_i^{\mathit{mis}}, Y_i^{\mathit{obs}}, X_i|\theta),
\end{equation}
which is proportional to a standard posterior distribution of $\theta$
had $\mathbf{D^{\mathit{mis}}}$ been~observed [\citet{jinprincipal2008}], further
motivating the strategy of first drawing $\mathbf{D^{\mathit{mis}}}$ and then
sampling from the posterior distribution of~$\theta$
conditional on complete compliance data. Posterior distributions of
the relevant quantities follow from specification of both $p(\theta)$
and the models defined in Section \ref{models}. We describe our prior
distributions for $\theta$ in Section~\ref{priors}.

\vspace*{-3pt}
\subsection{Models for principal strata and outcomes}\label{models}
To estimate the CACE, we further factor the joint distribution in (\ref
{joint}) as $f(\mathbf{Y|S,X})f(\mathbf{S|X})f(\mathbf{X})f(\mathbf
{Z})$, and specify models for $f(\mathbf{Y|S,X})$ and $f(\mathbf{S|X})$.
As the population consists of three underlying strata, we follow the
approach used in \citet{frangakisclustered2002} and \citet
{barnardprincipal2003}, whereby we
model $f(\mathbf{S|X})$ with two linked probit models, the first
modeling membership in the never-taker stratum and the second modeling
membership in the complier stratum conditional on exclusion from the
never-taker stratum. We parameterize these models as
%
\begin{eqnarray}\label{psimods}
\Psi_n(X_i, \beta) &= &P(S_i=n | X_i, \beta) = 1- \Phi(\beta_{00} + \beta
_{01}X_i), \nonumber\\
\Psi_c(X_i, \beta) &=& P(S_i=c | X_i, \beta) \nonumber\\[-11pt]\\[-11pt]
&=& \{1-\Psi_n(X_i, \beta)\}\{
1-\Phi(\beta_{10} + \beta_{11}X_i)\} \qquad \mbox{and}  \nonumber \\
\Psi_a(X_i, \beta) &=& P(S_i=a | X_i, \beta) = 1 - \Psi_n(X_i, \beta) -
\Psi_c(X_i, \beta), \nonumber
\end{eqnarray}
where $\beta= (\beta_{00}, \beta_{01}, \beta_{10}, \beta_{11})$ and
$\Phi$ is the standard normal cumulative distribution function. To
facilitate computation, we represent these models as arising from
underlying continuous random variables $S_i^n$ and $S_i^c$,
\begin{eqnarray}\label{latentmods}
S_i&=&n \quad\mathrm{if}\quad S_i^{n}=\beta_{00} + \beta_{01}X_i + V_i
\le0, \nonumber\\
S_i&=&c \quad\mathrm{if}\quad S_i^{n} > 0 \quad\mbox{and} \quad
S_i^{c} = \beta_{10} + \beta_{11}X_i + U_i \le0 \qquad \mathrm{and} \\
S_i&=&a \quad\mathrm{if}\quad S_i^{n}>0 \quad\mathrm{and} \quad
S_i^{c}>0, \nonumber
\end{eqnarray}
where the $V_i$ and $U_i$ are independently distributed as $N(0,1)$.

We illustrate the analysis with two different models for $f(\mathbf
{Y|S,X})$. The first model (Model A) entails a regression adjustment
for the key covariate's association with the outcome:
%
\begin{eqnarray}\label{gmodsA}
\qquad &&f\bigl(Y_i(z)|X_i, S_i=n\bigr)=g_n(Y_i|\alpha_0^n,\alpha_1^n, X_i, \sigma^2)
\sim N(\alpha_0^n + \alpha_1^nX_i, \sigma^2), \nonumber\\
\qquad &&f\bigl(Y_i(z)|X_i, S_i=a\bigr)=g_a(Y_i|\alpha_0^a,\alpha_1^a,X_i, \sigma^2)\sim
N(\alpha_0^a + \alpha_1^aX_i, \sigma^2) \quad\mathrm{and}\nonumber\\[-8pt]\\[-8pt]
\qquad &&f\bigl(Y_i(z)|X_i, S_i=c, Z_i=z\bigr)\nonumber \\
\qquad &&\qquad =g_{cz}(Y_i|\alpha_0^{cz},\alpha_1^{cz},X_i,
\sigma^2) \sim N(\alpha_0^{cz}+ \alpha_1^{cz}X_i, \sigma^2) \quad
\mbox{for }  z=0,1, \nonumber
\end{eqnarray}
implying the exclusion restriction and the assumption that $\mathit{GOHAI}$
outcomes are distributed with the same variance in each stratum and for
each treatment receipt. For comparison purposes, we also conduct the
analysis under another model (Model B) that does not explicitly
incorporate $X$ in the model for $Y(z)$, entailing the additional
assumption that $Y(z) \independent X |S$. That is, Model B incorporates
the restriction that $\alpha_1^n=\alpha_1^a=\alpha_1^{c0}=\alpha
_1^{c1}=0$, representing a ``standard'' unadjusted CACE analysis.

The observed-data likelihood reflecting the mixtures over the latent
$S_i$ can be written as
%
\begin{eqnarray}\label{obslike}
&&L_{obs}(\theta| \mathbf{Z, D^{\mathit{obs}}, Y^{\mathit{obs}}, X})\nonumber \\
&&\qquad  =
\prod_{Z_i=1, D_i^{\mathit{obs}}=0} \{\Psi_n(X_i, \beta) \cdot g_n(Y_i|\alpha
_0^n,\alpha_1^n, X_i, \sigma^2) \}\nonumber\\
&&\qquad \quad {}  \times\prod_{Z_i=0, D_i^{\mathit{obs}}=1} \{\Psi_a(X_i, \beta) \cdot g_a(Y_i|\alpha
_0^a,\alpha_1^a,X_i, \sigma^2) \}\nonumber \\
&&\qquad \quad {} \times\prod_{Z_i=0, D_i^{\mathit{obs}}=0} \{\Psi_n(X_i, \beta) \cdot g_n(Y_i|\alpha
_0^n,\alpha_1^n, X_i, \sigma^2)\\
&&\qquad \quad \hphantom{{} \times\prod_{Z_i=0, D_i^{\mathit{obs}}=0} \{}
{} + \Psi_c(X_i, \beta) \cdot
g_{c0}(Y_i|\alpha_0^{c0},\alpha_1^{c0},X_i, \sigma^2) \}\nonumber
\\
&&\qquad \quad {} \times\prod_{Z_i=1, D_i^{\mathit{obs}}=1} \{\Psi_a(X_i, \beta) \cdot g_a(Y_i|\alpha
_0^a,\alpha_1^a,X_i, \sigma^2) \nonumber \\
&&\qquad \quad \hphantom{{} \times\prod_{Z_i=0, D_i^{\mathit{obs}}=0} \{}
{} + \Psi_c(X_i, \beta) \cdot
g_{c1}(Y_i|\alpha_0^{c1},\alpha_1^{c1},X_i, \sigma^2) \} ,\nonumber
\end{eqnarray}
where $\theta=(\beta_{00}, \beta_{01}, \beta_{10}, \beta_{11}, \alpha
_0^n, \alpha_1^n, \alpha_0^a, \alpha_1^a, \alpha_0^{c0}, \alpha_1^{c0},
\alpha_0^{c1}, \alpha_1^{c1}, \sigma^2)$ and the product over $Z_i=z,
D^{\mathit{obs}}_i=d$ represents the product over all patients assigned
treatment $z$ who were observed to receive treatment $d$.

As a result of random assignment to treatment, the $Y_i^{\mathit{mis}}$ in the
stratum of compliers is sampled from the distribution $g_{c(1-Z_i)}$
and the CACE estimate is calculated as
\[
\mathit{CACE} = E[Y_i(1)-Y_i(0)|X_i, S_i=c]=\frac{1}{n_c} \sum_{S_i=c} \bigl(Y_i(1) -
Y_i(0)\bigr),
\]
where $n_c$ is the number of patients with $S_i=c$ at the current iteration.

\subsection{Sampling compliers within the compliance-predictive model}
Despite the existence of three underlying strata, the step of the Gibbs
sampler that determines patients' unknown compliance status does so via
Bernoulli distributions reflecting the fact that patients who received
the assigned treatment can belong to one of only two possible strata.
Owing to these underlying two-component mixtures, the probability at a
given iteration of the sampler that a patient with $Z_i=D^{\mathit{obs}}_i=z$
belongs to stratum of compliers is
%
\begin{eqnarray}\label{probcomplier}
&&P(S_i=c|X_i, Y_i^{\mathit{obs}}, D_i^{\mathit{obs}}, Z_i,\theta)\nonumber\\
&&\qquad =\Psi_c(X_i, \beta)
\cdot g_{cz}(Y_i|\alpha_0^{cz},\alpha_1^{cz},X_i,
\sigma^2)\nonumber\\[-8pt]\\[-8pt]
&&\qquad \quad {}/\bigl(\Psi
_c(X_i, \beta) \cdot g_{cz}(Y_i|\alpha_0^{cz},\alpha_1^{cz},X_i, \sigma
^2)\nonumber \\
&&\qquad \qquad
{} + \Psi_t(X_i, \beta) \cdot g_t(Y_i|\alpha_0^t,\alpha_1^t, X_i,
\sigma^2)\bigr) ,\nonumber
\end{eqnarray}
where $t=n$ if $z=0$ and $t=a$ if $z=1$. Examining these probabilities
makes clear that the relative impacts of $X_i$ and $Y^{\mathit{obs}}_i$ on (\ref
{probcomplier}) depend on the extent to which $X$ predicts stratum and
on the amount of overlap between the distributions $g_{cz}$ and~$g_{t}$.

\subsection{Additional model specifications and statistical computing
details}\label{priors}
We treat the elements of $\theta$ to be a priori independent,
using conditionally-con\-jugate normal distributions for the $\beta,
\alpha^n_0, \alpha^n_1, \alpha^a_0, \alpha^a_1, \alpha^{cz}_0, \alpha
^{cz}_1$, and a conditio\-nally-conjugate gamma distribution for the
precision parameter $\frac{1}{\sigma^2}$. The distributions for $(\alpha
_0^n, \alpha_0^a, \alpha_0^{cz})$ are centered at the overall sample
mean $\mathit{GOHAI}$ with variances of 100, and the distributions for $(\alpha
_1^n, \alpha_1^a,\alpha_1^{cz})$ are centered at~0 with variances of
100. The prior distribution for the precision parameter is gamma with
shape and scale parameter set to 0.01. Prior distributions for the
elements of $\beta$ are centered at 0 with variance 5.

After a burn-in of 5,000 iterations, each chain is run for 5,000
additional iterations, saving every $10$th sample. For each model,
three chains are run from different starting values, and the potential
scale-reduction statistics [\citet{gelmaninference1992}; \citet{gelmanbayesian2004}]
are calculated for each parameter to assess convergence. All
parameters in all models had potential scale-reduction statistics less
than or equal to 1.06, suggesting satisfactory convergence. For each
model, the three chains are combined to calculate posterior estimates.

\section{Illustration of the potential for Bias in the CACE using
simulated data}\label{sims}
To illustrate that the compliance-predictive model can imply complier
treatment groups with different characteristics and to illustrate our
graphical diagnostic, we examine in detail a simulated scenario where
$X$ is predictive of stratum membership and the true $\mathit{CACE}=0$. Details of
this simulation and a broader simulation study appear in a
supplementary web appendix [\citet{ziglersupplement2011}].

\begin{figure}[b]

\includegraphics{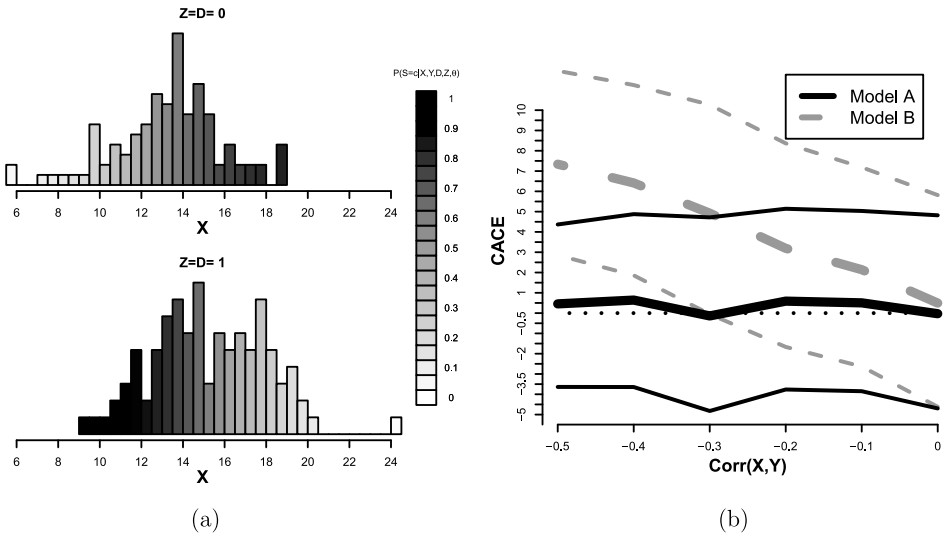}

\caption{Results from simulated data sets
where $X$ predicts stratum membership. As the procedure selects compliers from
opposite ends of the severity distribution (\textup{a}), estimates
of the CACE can become particularly susceptible to model
misspecification (\textup{b}). Thick lines in (\textup{b}) are
posterior means and thin lines are 95\% posterior intervals. For each
value of $\operatorname{Corr}(X,Y)$, posterior summaries are averaged over 50 Monte
Carlo simulations. All simulations have $\mathit{CACE}=0$ (horizontal dotted
line). (\textup{a})~Observed $\mathit{SEV}$ distributions shaded corresponding to
$P(S_i=c|X_i,Y_i^{\mathit{obs}}, D_i^{\mathit{obs}}, Z_i,\theta)$. (\textup{b}) Posterior CACE estimates under Models A and B.}\label{simfigs}
\end{figure}

To investigate the relationships between $X$ and stratum membership
under Model A, we examine posterior-predictive distributions of the
probabilities in (\ref{probcomplier}) for a hypothetical group of
patients having an $X$ distribution mirroring that in the observed
data. Figure \ref{simfigs}(a) displays, for $z=0,1$, histograms of the
observed $X$ distributions in patients with $Z_i=D_i^{\mathit{obs}}=z$,
with histogram bars shaded according to the mean posterior-predictive
probability of membership in the complier stratum for a value of $X$ at
that point of the histogram and for $Y_i^{\mathit{obs}}$ equal to the
mean value observed in patients with $Z_i=D_i^{\mathit{obs}}=z$. Note
the different shading patterns in the two histograms. For
$Z_i=D_i^{\mathit{obs}}=0$, histogram bars are darker as $X$ increases
(more severely injured patients are more likely compliers), while for
$Z_i=D_i^{\mathit{obs}}=1$, histogram bars are lighter as $X$ increases
(more severely injured patients are less likely compliers). For
example, note that patients with $X$ in the range $[9, 12]$ in the
$Z_i=D_i^{\mathit{obs}}=1$ group have probability of membership in the
complier stratum near $1.0$, while patients with the same range of
$\mathit{SEV}$ in the $Z_i=D_i^{\mathit{obs}}=0$ group have probability of
membership in the complier stratum in the range $[0.1-0.4]$. The
implication for estimates of the CACE is that over the course of the
sampler, patients in the observed mixture of compliers and
always-takers ($Z_i=D_i^{\mathit{obs}}=1$) with lower $X$ will more
often contribute to the CACE than patients with comparable $X$ in the
observed mixture of compliers and never-takers
($Z_i=D_i^{\mathit{obs}}=0$). The opposite sampling disparity holds for
patients with higher~$X$. If $X$ is also related to the primary
outcome, $Y$, a situation such as that depicted in Figure \ref{simfigs}(a)
leaves estimates of the CACE particularly vulnerable to misspecified
models that incorrectly extrapolate to areas of the $X$ distribution
where there is limited data. For example, estimation of a typical
unadjusted CACE (Model~B) would represent one such misspecified model,
and could lead to vastly different estimates of the CACE. To illustrate
this point, Figure \ref{simfigs}(b) displays posterior estimates of the CACE
using both Model A and Model B in scenarios where $X$ is related to
stratum membership and with varying magnitudes of the relationship
between $X$ and $Y$. We see that under Model B, the imbalanced sampling
of compliers evident from Figure \ref{simfigs}(a) leads to bias in the
estimated CACE that is increasing in $|\operatorname{Corr}(X,Y)|$, providing
misleading results even when the association between $X$ and $Y$ is
modest and in some cases estimating a significant treatment effect when
there in fact is none. The same bias is not depicted under Model A
because even though there is limited data on comparable compliers in
some areas of the $X$ distribution, extrapolation of Model A to these
areas of the distribution correctly reflects the underlying
relationship; that is, there is no model misspecification. The
supplementary web appendix [\citet{ziglersupplement2011}]
considers simulations under a broader range of relationships between
$X$ and both stratum membership and $Y$ and further indicates the
potential for bias in the CACE when using a compliance-predictive
covariate.

\section{Using $\mathit{SEV}$ to predict principal strata in the motivating
oral-surgery study}\label{dentdat}

As described in Table \ref{compsum}, patients in the oral-surgery
example who had assignment to MMF overruled (known always-takers) had
higher average $X$ than the rest of the sample, patients who had
assignment to RIF overruled (known never-takers) had lower average $X$
than the rest of the sample, and there was a relatively high estimated
proportion of never-takers (50.0\%) and a relatively low proportion of
always-takers (14.5\%). The oral-surgery example had missing $Y$ for a
substantial proportion of the patients. Based on observed data, the
nonresponse rates were 48.4\%, 46.2\% in the $Z=0,1$ arms,
respectively, and 55.6\%, 45.0\% in the observed always-takers,
never-takers, respectively. To prevent complication of our illustrative
goal, we assume in the models for the oral-surgery data that (1) the
$S_i$ are independent of the missing indicator and (2) the missing $Y$
are latently ignorable conditional on $S_i$ and $Z_i$
[\citet{frangakisaddressing1999}]. The implication of these assumptions for the computation
is that missing $Y$ are drawn at each iteration from the distribution
for patients' current stratum membership conditional on current values
of the parameters. Furthermore, the small number of observed $Y$ values
precludes useful estimation of all of the $\alpha$ parameters in (\ref
{gmodsA}), leading us to alter Model A to Model A$^*$ that includes the
constraint that $\alpha_1^n = \alpha_1^a=\alpha_1^{c0}=\alpha_1^{c1}$.

\begin{figure}

\includegraphics{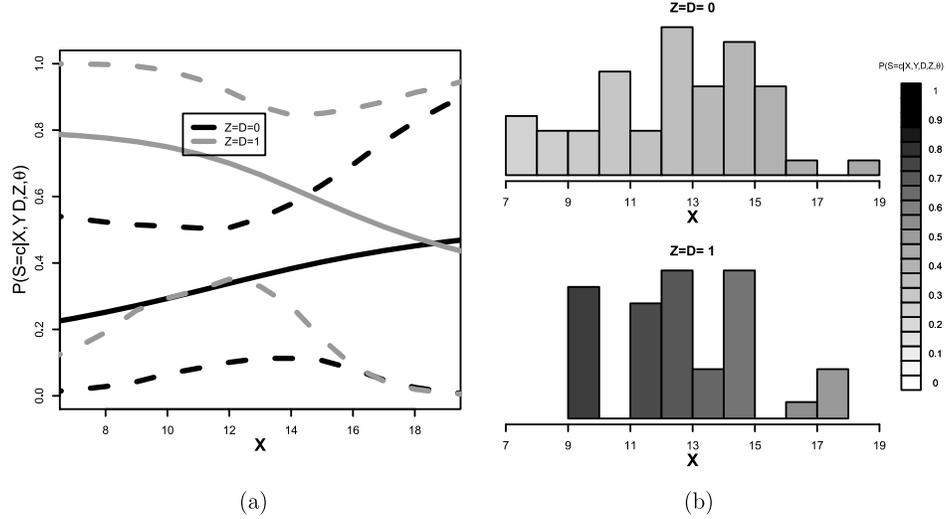}

\caption{Posterior-predicted probabilities of membership in the
complier stratum for hypothetical patients with $Z_i=D_i^{\mathit{obs}}=z$ in
the oral-surgery study under Model A$^*$, $z=0,1$. (\textup{a}) Posterior mean (solid) and 95\% intervals (dashed) for
$P(S_i=c|X_i,Y_i^{\mathit{obs}},D_i^{\mathit{obs}}, Z_i, \theta)$. (\textup{b}) Observed $\mathit{SEV}$ distributions shaded corresponding to
$P(S_i=c|X_i,Y_i^{\mathit{obs}}, D_i^{\mathit{obs}}, Z_i,\theta)$.}\label{postprobsdent}
\vspace*{-3pt}
\end{figure}

The observed relationship between $X$ and membership in the never-taker
and always-taker strata (Table \ref{compsum}) prompts examination of
the probabilities of selection into the stratum of compliers within the
compliance-predictive model. Figure \ref{postprobsdent}(a) shows the
posterior predictive distributions of the Ber\-noulli probabilities in
(\ref{probcomplier}) for hypothetical patients with $Y$ equal to the
observed sample mean, $X$ across the range observed in the data, and
$Z_i=D_i^{\mathit{obs}}=z$ for $z=0,1$. The unequal proportions of underlying
strata are reflected in this figure by the fact that the probability of
being sampled as a~complier is\vadjust{\eject} consistently higher for the patients
with $Z_i=D_i^{\mathit{obs}}=1$ than for those in the other treatment arm; the
low estimated proportion of always-takers ($9/62=0.14$) implies that most
patients with $Z_i=D_i^{\mathit{obs}}=1$ belong to the stratum of compliers.

\begin{table}[b]
\tablewidth=330pt
\caption{Posterior estimates of the CACE in the motivating oral-surgery
study using a model without the compliance-predictive covariate and
using three compliance-predictive strategies}\label{CACEdenttab}
\begin{tabular*}{\tablewidth}{@{\extracolsep{\fill}}lccd{3.1}c@{}}
\hline
\textbf{Modeling strategy} & \textbf{Posterior mean} & \textbf{SD} &\multicolumn{1}{c}{\textbf{2.5\%}} & \textbf{97.5\%} \\
\hline
Compliance-predictive Model A$^*$&2.65&7.3&-8.9&20.9 \\
Compliance-predictive Model B&0.17&7.0&-13.0&16.0\\
Compliance-predictive Model C$^*$&1.95&6.7& -9.6 &18.9 \\
Model without $\mathit{SEV}$&0.74&6.4&-11.0&13.6\\
\hline
\end{tabular*}
\end{table}

The wide spread of the posterior predictive distributions in Figure \ref{postprobsdent}(a) suggests that $X$ has limited utility for
identifying which patients are compliers, but there is some indication
that the relationship between $X$ and the probability of membership in
the complier stratum is slightly different at the high end of the $X$
distribution depending on the value of $Z_i$ and $D_i^{\mathit{obs}}$. To assess
the potential for these relationships to affect the sampling of
compliers, we examine in Figure \ref{postprobsdent}(b) the
posterior-predictive probabilities of membership in the complier
stratum for hypothetical patients with $X$ distributions identical to
those observed in the sample with $Z_i=D_i^{\mathit{obs}}=z$ and with $Y$ equal
to the mean value observed in patients with $Z_i=D_i^{\mathit{obs}}=z$, for
$z=0,1$. This illustration provides limited evidence that patients with
different values of $Z_i$ and $D_i^{\mathit{obs}}$ are sampled as compliers from
different areas of their respective $X$ distributions. There is a
slight positive association between $X$ and membership in the complier
stratum in the $Z_i=D_i^{\mathit{obs}}=0$ patients (evidenced by the darkening
of the histogram bars as $X$ increases) that differs from the negative
association in the $Z_i=D_i^{\mathit{obs}}=1$ patients (evidenced by the
lightening of the histogram bars as $X$ increases), but the amount of
uncertainty in these posterior probabilities likely precludes any
serious effect on the estimated CACE.

Overall, the information in Figure \ref{postprobsdent} does not provide
any strong indication that the compliance-predictive model estimates a
CACE calculated from compliers with different injury characteristics in
the two treatment groups. To explore the sensitivity to alternative
models for stratum membership, we adapt Model A$^*$ to replace the
probit models in (\ref{psimods}) with a~multinomial logit model along
the lines of that used in \citet{hiranoassessing2000}, and refer to this
as Model C$^*$. Using Model C$^*$, figures analogous to Figure~\ref{postprobsdent}(a) and~(b)
appear largely
indistinguishable from those under Model~A$^*$ and are not pictured.
Table \ref{CACEdenttab} summarizes posterior CACE estimates from a~compliance-predictive
analysis under Model A$^*$, Model B and Model C$^*$, as well as from an analysis
following \citet{imbensbayesian1997} that does not explicitly use the
$\mathit{SEV}$ covariate at all and places a noninformative
conditionally-conjugate Dirichlet prior distribution on the population
proportions of principal strata. None of these models provide
evidence of a~treatment effect, and all three compliance-predictive
models offer slightly decreased precision, most likely due to the lack
of information contained in the $\mathit{SEV}$ covariate regarding stratum
membership and the inclusion of extraneous model parameters.

\section{Discussion}
Using covariates to model membership in latent principal strata has
many advantages in estimating the CACE. We provide a detailed
illustration of the subtlety involved in using a key covariate when
noncompliance exists in both treatment arms. In particular, we show
that when a~covariate is related to stratum membership, a
joint-estimation method can imply treatment groups in the latent
stratum of compliers with different covariate characteristics. The
resulting danger of comparing compliers with different characteristics
can be alleviated with modeling assumptions that correctly extrapolate
the treatment effect to areas of the covariate distribution where
compliers are not estimated to exist in both treatment arms. However,
this differential sampling of patients into the complier stratum poses
a serious threat to the CACE under model misspecification, including
calculation of the standard unadjusted CACE when a covariate predicts
stratum membership. We propose simple graphical posterior checks that
indicate the extent to which the estimated CACE relies on compliers
that have different covariate characteristics, potentially
characterizing the danger for model misspecification to bias the
estimated CACE.

Our aim is not to discourage the use of covariates that are predictive
of latent stratum membership but rather to shed light on the subtleties
involved and to provide guidance on how to detect whether a
compliance-predictive model endangers estimates of the CACE. Our
motivating oral-surgery example is somewhat unique in its availability
of a key covariate that was thought to influence the treatment
received, but the possibility of covariates relating to stratum
membership can arise elsewhere, as with the randomized encouragement
design considered in \citet{hiranoassessing2000} where age and presence
of chronic obstructive pulmonary disease (COPD) were thought to
influence whether patients were in the underlying stratum of
individuals who would always receive a flu vaccination regardless of
random encouragement to do so. The authors of that work include
compliance-predictive models to relax exclusion restrictions and
provide posterior estimates of model parameters suggestive of a
different relationship between age and COPD and the probability of
membership in the complier stratum depending on the values of~$Z_i$ and
$D_i^{\mathit{obs}}$. Whether their model tended to consider a complier stratum
consisting of younger patients without COPD in the $Z=1$ arm and older
patients with COPD in the $Z=0$ arm could be assessed by examining
posterior probabilities of stratum membership across the observed
ranges of these covariates.

We present models that do and do not adjust the CACE for levels of the
key covariate. We frame the choice not to model $Y$ conditional on
both~$S$ and $X$ (as in Model B) as a form of model misspecification, but in
real applications researchers are confronted with the decision to
calculate the familiar unadjusted CACE or to specify a more detailed
model for $f(\mathbf{Y|S,X})$ and estimate an adjusted CACE. We show
that when a covariate is used to model stratum membership, estimation
of the unadjusted CACE can produce biased results. Thus, we recommend
that the CACE be adjusted for any covariates used to model stratum
membership, which is contrary to previous recommendations that
stratum-predictive covariates need not be included in models for
outcomes within strata [\citet{gallopmediation2009}]. Furthermore,
specification of a more detailed model for
$f(\mathbf{Y|S,X})$ does not guarantee correctness, and we provide a
framework to assess whether model misspecification poses a particular
danger to estimation of a covariate-adjusted CACE that can depend on
areas of the covariate distribution where there is limited
data.\looseness=1

The core features of the scenario presented here, namely, that a
complian\-ce-predictive model must respect the presence of three
underlying strata while a patient of unknown stratum can belong to one
of only two strata, can have conflicting impacts. One way to
characterize these issues is to view modeling membership in the
complier stratum not as selection of compliers but rather as a process
for selection of ``nonnoncompliers'' from both treatment arms since,
no matter how predictive, the compliance-predictive feature is anchored
to observed information on always-takers and never-takers and can only
indirectly model membership in the stratum of primary interest. Some
applications focus on treatment effects within principal strata
analogous to always-takers [\citet{hudgensanalysis2003};
\citet{gilbertsensitivity2003};
\citet{shepherdsensitivity2006}; \citet{hudgenscausal2006};
\citet{royprincipal2008}] and are less susceptible to the type of
bias depicted here because, as in settings where noncompliance exists
in only one treatment arm, the data provide direct evidence on the
relationship between covariates and the stratum of primary interest.

We have characterized scenarios that lend themselves to the use of
a~com\-pliance-predictive covariate but leave an opening for bias in the
estimation of the CACE. Such scenarios warrant careful model checking;
in Sections \ref{sims} and \ref{dentdat} we propose steps to
investigate the potential for bias. Future research on methods that use
stratum-predictive covariates to estimate the CACE when the data
consist of three underlying strata would prove valuable in settings
where it is appealing to use covariates to aid identifiability or
improve precision of causal estimates.

\begin{supplement}
\stitle{Simulation study}
\slink[doi]{10.1214/11-AOAS477SUPP}
\slink[url]{http://lib.stat.cmu.edu/aoas/477/supplement.pdf}
\sdatatype{.pdf}
\sdescription{A detailed exposition of the potential for bias using a
richer set of simulations.}
\end{supplement}


%

\printaddresses


\begin{thebibliography}{36}

%
%
\bibitem[\protect\citeauthoryear{Angrist, Imbens and
Rubin}{1996}]{angristidentification1996}
\begin{barticle}[author]
\bauthor{\bsnm{Angrist},~\bfnm{J.~D.}\binits{J.~D.}},
\bauthor{\bsnm{Imbens},~\bfnm{G.~W.}\binits{G.~W.}} \AND
\bauthor{\bsnm{Rubin},~\bfnm{D.~B.}\binits{D.~B.}}
(\byear{1996}).
\btitle{Identification of causal effects using instrumental variables}.
\bjournal{J. Amer. Statist. Assoc.}
\bvolume{91}
\bpages{444--455}.
\end{barticle}
\endbibitem

%
%
\bibitem[\protect\citeauthoryear{Atchison}{1997}]{atchisongeneral1997}
\begin{bmisc}[author]
\bauthor{\bsnm{Atchison},~\bfnm{Kathryn}\binits{K.}}
(\byear{1997}).
\bhowpublished{The general oral health assessment index.
In \textit{Measuring Oral Health and Quality of Life}
71--80 Univ. North Carolina, Chapel Hill, NC}.
\end{bmisc}
\endbibitem

%
%
\bibitem[\protect\citeauthoryear{Barnard et~al.}{2003}]{barnardprincipal2003}
\begin{barticle}[author]
\bauthor{\bsnm{Barnard},~\bfnm{J.}\binits{J.}},
\bauthor{\bsnm{Frangakis},~\bfnm{C.~E}\binits{C.~E.}},
\bauthor{\bsnm{Hill},~\bfnm{J.~L}\binits{J.~L.}} \AND
\bauthor{\bsnm{Rubin},~\bfnm{D.~B}\binits{D.~B.}}
(\byear{2003}).
\btitle{Principal stratification approach to broken randomized
experiments: A
case study of school choice Vouchers in New York city.}
\bjournal{J. Amer. Statist. Assoc.}
\bvolume{98}
\bpages{299--323}.
\end{barticle}
\MR{1995712}
\endbibitem

%
%
\bibitem[\protect\citeauthoryear{de~Finetti}{1974}]{definettitheory1974}
\begin{bbook}[author]
\bauthor{\bparticle{de} \bsnm{Finetti},~\bfnm{B.}\binits{B.}}
(\byear{1974}).
\btitle{Theory of Probability: A Critical Introductory Treatment}.
\bpublisher{Wiley}, \baddress{New York}.
\end{bbook}
\endbibitem

%
%
\bibitem[\protect\citeauthoryear{Follmann}{2000}]{follmanneffect2000}
\begin{barticle}[author]
\bauthor{\bsnm{Follmann},~\bfnm{D.~A}\binits{D.~A.}}
(\byear{2000}).
\btitle{On the effect of treatment among would-be treatment compliers: An
analysis of the Multiple Risk Factor Intervention Trial.}
\bjournal{J. Amer. Statist. Assoc.}
\bvolume{95}
\bpages{1101--1109}.
\end{barticle}
\MR{1821718}
\endbibitem

%
%
\bibitem[\protect\citeauthoryear{Frangakis and
Rubin}{1999}]{frangakisaddressing1999}
\begin{barticle}[author]
\bauthor{\bsnm{Frangakis},~\bfnm{Constantine~E.}\binits{C.~E.}} \AND
\bauthor{\bsnm{Rubin},~\bfnm{Donald~B.}\binits{D.~B.}}
(\byear{1999}).
\btitle{Addressing complications of intention-to-treat analysis in the combined
presence of all-or-none treatment-noncompliance and subsequent missing
outcomes}.
\bjournal{Biometrika}
\bvolume{86}
\bpages{365--379}.
\end{barticle}
\MR{1705410}
\endbibitem

%
%
\bibitem[\protect\citeauthoryear{Frangakis and
Rubin}{2002}]{frangakisprincipal2002}
\begin{barticle}[author]
\bauthor{\bsnm{Frangakis},~\bfnm{Constantine~E.}\binits{C.~E.}} \AND
\bauthor{\bsnm{Rubin},~\bfnm{Donald~B.}\binits{D.~B.}}
(\byear{2002}).
\btitle{Principal stratification in causal inference}.
\bjournal{Biometrics}
\bvolume{58}
\bpages{21--29}.
\end{barticle}
\MR{1891039}
\endbibitem

%
%
\bibitem[\protect\citeauthoryear{Frangakis, Rubin and
Zhou}{2002}]{frangakisclustered2002}
\begin{barticle}[author]
\bauthor{\bsnm{Frangakis},~\bfnm{Constantine~E.}\binits{C.~E.}},
\bauthor{\bsnm{Rubin},~\bfnm{Donald~B.}\binits{D.~B.}} \AND
\bauthor{\bsnm{Zhou},~\bfnm{{Xiao-Hua}}\binits{X.}}
(\byear{2002}).
\btitle{Clustered encouragement designs with individual noncompliance:
{{B}ayesian} inference with randomization, and application to advance
directive forms}.
\bjournal{Biostatistics}
\bvolume{3}
\bpages{147--164}.
\end{barticle}
\endbibitem

%
%
\bibitem[\protect\citeauthoryear{Gallop et~al.}{2009}]{gallopmediation2009}
\begin{barticle}[author]
\bauthor{\bsnm{Gallop},~\bfnm{Robert}\binits{R.}},
\bauthor{\bsnm{Small},~\bfnm{Dylan~S.}\binits{D.~S.}},
\bauthor{\bsnm{Lin},~\bfnm{Julia~Y.}\binits{J.~Y.}},
\bauthor{\bsnm{Elliott},~\bfnm{Michael~R.}\binits{M.~R.}},
\bauthor{\bsnm{Joffe},~\bfnm{Marshall}\binits{M.}} \AND\bauthor{\bsnm{{{Ten}
Have}},~\bfnm{Thomas~R.}\binits{T.~R.}}
(\byear{2009}).
\btitle{Mediation analysis with principal stratification}.
\bjournal{Stat. Med.}
\bvolume{28}
\bpages{1108--1130}.
\end{barticle}
\MR{2662200}
\endbibitem

%
%
\bibitem[\protect\citeauthoryear{Gelfand and
Smith}{1990}]{gelfandsampling-based1990}
\begin{barticle}[author]
\bauthor{\bsnm{Gelfand},~\bfnm{Alan~E.}\binits{A.~E.}} \AND
\bauthor{\bsnm{Smith},~\bfnm{Adrian F.~M.}\binits{A.~F.~M.}}
(\byear{1990}).
\btitle{Sampling-based approaches to calculating marginal densities}.
\bjournal{J. Amer. Statist. Assoc.}
\bvolume{85}
\bpages{398--409}.
\end{barticle}
\MR{1141740}
\endbibitem

%
%
\bibitem[\protect\citeauthoryear{Gelman and Rubin}{1992}]{gelmaninference1992}
\begin{barticle}[author]
\bauthor{\bsnm{Gelman},~\bfnm{A.}\binits{A.}} \AND
\bauthor{\bsnm{Rubin},~\bfnm{D.~B}\binits{D.~B.}}
(\byear{1992}).
\btitle{Inference from iterative simulation using multiple sequences}.
\bjournal{Statist. Sci.}
\bvolume{7}
\bpages{457--472}.
\end{barticle}
\endbibitem

%
%
\bibitem[\protect\citeauthoryear{Gelman et~al.}{2004}]{gelmanbayesian2004}
\begin{bbook}[author]
\bauthor{\bsnm{Gelman},~\bfnm{A.}\binits{A.}},
\bauthor{\bsnm{Carlin},~\bfnm{J.~B}\binits{J.~B.}},
\bauthor{\bsnm{Stern},~\bfnm{H.~S}\binits{H.~S.}} \AND
\bauthor{\bsnm{Rubin},~\bfnm{D.~B}\binits{D.~B.}}
(\byear{2004}).
\btitle{Bayesian Data Analysis}, \bedition{2nd} ed.
\bpublisher{Chapman \& Hall/CRC}, \baddress{Boca Raton, FL}.
\end{bbook}
\MR{2027492}
\endbibitem

%
%
\bibitem[\protect\citeauthoryear{Geman and Geman}{1984}]{gemanstochastic1984}
\begin{barticle}[author]
\bauthor{\bsnm{Geman},~\bfnm{S.}\binits{S.}} \AND
\bauthor{\bsnm{Geman},~\bfnm{D.}\binits{D.}}
(\byear{1984}).
\btitle{Stochastic relaxation, {{G}ibbs} distributions, and the {{B}ayesian}
restoration of images}.
\bjournal{{IEEE} Transactions on Pattern Analysis and Machine Intelligence}
\bvolume{6}
\bpages{721--741}.
\end{barticle}
\endbibitem

%
%
\bibitem[\protect\citeauthoryear{Gilbert, Bosch and
Hudgens}{2003}]{gilbertsensitivity2003}
\begin{barticle}[author]
\bauthor{\bsnm{Gilbert},~\bfnm{P.~B.}\binits{P.~B.}},
\bauthor{\bsnm{Bosch},~\bfnm{R.~J.}\binits{R.~J.}} \AND
\bauthor{\bsnm{Hudgens},~\bfnm{M.~G.}\binits{M.~G.}}
(\byear{2003}).
\btitle{Sensitivity analysis for the assessment of causal vaccine
effects on
viral load in {HIV} vaccine trials}.
\bjournal{Biometrics}
\bvolume{59}
\bpages{531--541}.
\end{barticle}
\MR{2004258}
\endbibitem

%
%
\bibitem[\protect\citeauthoryear{Griffin, {McCaffrey} and
Morral}{2008}]{griffinapplication2008}
\begin{barticle}[author]
\bauthor{\bsnm{Griffin},~\bfnm{Beth~Ann}\binits{B.~A.}},
\bauthor{\bsnm{{McCaffrey}},~\bfnm{Daniel~F}\binits{D.~F.}} \AND
\bauthor{\bsnm{Morral},~\bfnm{Andrew~R}\binits{A.~R.}}
(\byear{2008}).
\btitle{An application of principal stratification to control for
institutionalization at follow-up in studies of substance abuse treatment
programs}.
\bjournal{Ann. Appl. Stat.}
\bvolume{2}
\bpages{1034--1055}.
\end{barticle}
\MR{2516803}
\endbibitem

%
%
\bibitem[\protect\citeauthoryear{Hill, {Brooks-Gunn} and
Waldfogel}{2003}]{hillsustained2003}
\begin{barticle}[author]
\bauthor{\bsnm{Hill},~\bfnm{J.~L}\binits{J.~L.}},
\bauthor{\bsnm{{Brooks-Gunn}},~\bfnm{J.}\binits{J.}} \AND
\bauthor{\bsnm{Waldfogel},~\bfnm{J.}\binits{J.}}
(\byear{2003}).
\btitle{Sustained effects of high participation in an early
intervention for
low-birth-weight premature infants}.
\bjournal{Developmental Psychology}
\bvolume{39}
\bpages{730--744}.
\end{barticle}
\endbibitem

%
%
\bibitem[\protect\citeauthoryear{Hirano et~al.}{2000}]{hiranoassessing2000}
\begin{barticle}[author]
\bauthor{\bsnm{Hirano},~\bfnm{Keisuke}\binits{K.}},
\bauthor{\bsnm{Imbens},~\bfnm{Guido~W.}\binits{G.~W.}},
\bauthor{\bsnm{Rubin},~\bfnm{Donald~B.}\binits{D.~B.}} \AND
\bauthor{\bsnm{Zhou},~\bfnm{{Xiao-Hua}}\binits{X.}}
(\byear{2000}).
\btitle{Assessing the effect of an influenza vaccine in an encouragement
design}.
\bjournal{Biostatistics}
\bvolume{1}
\bpages{69--88}.
\end{barticle}
\endbibitem

%
%
\bibitem[\protect\citeauthoryear{Holland}{1986}]{hollandstatistics1986}
\begin{barticle}[author]
\bauthor{\bsnm{Holland},~\bfnm{Paul~W.}\binits{P.~W.}}
(\byear{1986}).
\btitle{Statistics and causal inference}.
\bjournal{J. Amer. Statist. Assoc.}
\bvolume{81}
\bpages{945--960}.
\bnote{With discussion and a reply by the author}.
\end{barticle}
\MR{0867618}
\endbibitem

%
%
\bibitem[\protect\citeauthoryear{Hudgens and
Halloran}{2006}]{hudgenscausal2006}
\begin{barticle}[author]
\bauthor{\bsnm{Hudgens},~\bfnm{Michael~G.}\binits{M.~G.}} \AND
\bauthor{\bsnm{Halloran},~\bfnm{M.~Elizabeth}\binits{M.~E.}}
(\byear{2006}).
\btitle{Causal vaccine effects on binary postinfection outcomes}.
\bjournal{J. Amer. Statist. Assoc.}
\bvolume{101}
\bpages{51--64}.
\end{barticle}
\MR{2252433}
\endbibitem

%
%
\bibitem[\protect\citeauthoryear{Hudgens, Hoering and
Self}{2003}]{hudgensanalysis2003}
\begin{barticle}[author]
\bauthor{\bsnm{Hudgens},~\bfnm{Michael~G.}\binits{M.~G.}},
\bauthor{\bsnm{Hoering},~\bfnm{Antje}\binits{A.}} \AND
\bauthor{\bsnm{Self},~\bfnm{Steven~G.}\binits{S.~G.}}
(\byear{2003}).
\btitle{On the analysis of viral load endpoints in {HIV} vaccine trials}.
\bjournal{Stat. Med.}
\bvolume{22}
\bpages{2281--2298}.
\end{barticle}
\endbibitem

%
%
\bibitem[\protect\citeauthoryear{Imbens and
Angrist}{1994}]{imbensidentification1994}
\begin{barticle}[author]
\bauthor{\bsnm{Imbens},~\bfnm{Guido~W.}\binits{G.~W.}} \AND
\bauthor{\bsnm{Angrist},~\bfnm{Joshua~D.}\binits{J.~D.}}
(\byear{1994}).
\btitle{Identification and estimation of local average treatment effects}.
\bjournal{Econometrica}
\bvolume{62}
\bpages{467--467}.
\end{barticle}
\endbibitem

%
%
\bibitem[\protect\citeauthoryear{Imbens and Rubin}{1997}]{imbensbayesian1997}
\begin{barticle}[author]
\bauthor{\bsnm{Imbens},~\bfnm{Guido~W.}\binits{G.~W.}} \AND
\bauthor{\bsnm{Rubin},~\bfnm{Donald~B.}\binits{D.~B.}}
(\byear{1997}).
\btitle{Bayesian inference for causal effects in randomized experiments with
noncompliance}.
\bjournal{Ann. Statist.}
\bvolume{25}
\bpages{305--327}.
\end{barticle}
\MR{1429927}
\endbibitem

%
%
\bibitem[\protect\citeauthoryear{Jin and Rubin}{2008}]{jinprincipal2008}
\begin{barticle}[author]
\bauthor{\bsnm{Jin},~\bfnm{H.}\binits{H.}} \AND
\bauthor{\bsnm{Rubin},~\bfnm{D.~B.}\binits{D.~B.}}
(\byear{2008}).
\btitle{Principal stratification for causal inference with extended partial
compliance}.
\bjournal{J. Amer. Statist. Assoc.}
\bvolume{103}
\bpages{101--111}.
\end{barticle}
\MR{2463484}
\endbibitem

%
%
\bibitem[\protect\citeauthoryear{Jo and Stuart}{2009}]{jouse2009}
\begin{barticle}[author]
\bauthor{\bsnm{Jo},~\bfnm{Booil}\binits{B.}} \AND
\bauthor{\bsnm{Stuart},~\bfnm{Elizabeth~A}\binits{E.~A.}}
(\byear{2009}).
\btitle{On the use of propensity scores in principal causal effect estimation}.
\bjournal{Stat. Med.}
\bvolume{28}
\bpages{2857--2875}.
\end{barticle}
\MR{2750169}
\endbibitem

%
%
\bibitem[\protect\citeauthoryear{Joffe, Small and
Hsu}{2007}]{joffedefining2007}
\begin{barticle}[author]
\bauthor{\bsnm{Joffe},~\bfnm{M.~M}\binits{M.~M.}},
\bauthor{\bsnm{Small},~\bfnm{D.}\binits{D.}} \AND
\bauthor{\bsnm{Hsu},~\bfnm{C.~Y}\binits{C.~Y.}}
(\byear{2007}).
\btitle{Defining and estimating intervention effects for groups that will
develop an auxiliary outcome}.
\bjournal{Statist. Sci.}
\bvolume{22}
\bpages{74--97}.
\end{barticle}
\MR{2408662}
\endbibitem

%
%
\bibitem[\protect\citeauthoryear{Joffe, Ten Have and
Brensinger}{2003}]{joffecompliance2003}
\begin{barticle}[author]
\bauthor{\bsnm{Joffe},~\bfnm{M.~M}\binits{M.~M.}}, \bauthor{\bsnm{{{Ten}
Have}},~\bfnm{T.~R.}\binits{T.~R.}} \AND
\bauthor{\bsnm{Brensinger},~\bfnm{C.}\binits{C.}}
(\byear{2003}).
\btitle{The compliance score as a regressor in randomized trials}.
\bjournal{Biostatistics}
\bvolume{4}
\bpages{327--340}.
\end{barticle}
\endbibitem

%
%
\bibitem[\protect\citeauthoryear{{McClellan}, {McNeil} and
Newhouse}{1994}]{mcclellandoes1994}
\begin{barticle}[author]
\bauthor{\bsnm{{McClellan}},~\bfnm{M.}\binits{M.}},
\bauthor{\bsnm{{McNeil}},~\bfnm{B.~J.}\binits{B.~J.}} \AND
\bauthor{\bsnm{Newhouse},~\bfnm{J.~P.}\binits{J.~P.}}
(\byear{1994}).
\btitle{Does more intensive treatment of acute myocardial infarction in the
elderly reduce mortality? {{A}nalysis} using instrumental variables}.
\bjournal{Journal of the American Medical Association}
\bvolume{272}
\bpages{859--866}.
\end{barticle}
\endbibitem

%
%
\bibitem[\protect\citeauthoryear{Rosenbaum and
Rubin}{1983}]{rosenbaumcentral1983}
\begin{barticle}[author]
\bauthor{\bsnm{Rosenbaum},~\bfnm{Paul~R.}\binits{P.~R.}} \AND
\bauthor{\bsnm{Rubin},~\bfnm{Donald~B.}\binits{D.~B.}}
(\byear{1983}).
\btitle{The central role of the propensity score in observational
studies for
causal effects}.
\bjournal{Biometrika}
\bvolume{70}
\bpages{41--55}.
\end{barticle}
\MR{0742974}
\endbibitem

%
%
\bibitem[\protect\citeauthoryear{Roy, Hogan and
Marcus}{2008}]{royprincipal2008}
\begin{barticle}[author]
\bauthor{\bsnm{Roy},~\bfnm{Jason}\binits{J.}},
\bauthor{\bsnm{Hogan},~\bfnm{Joseph~W}\binits{J.~W.}} \AND
\bauthor{\bsnm{Marcus},~\bfnm{Bess~H}\binits{B.~H.}}
(\byear{2008}).
\btitle{Principal stratification with predictors of compliance for randomized
trials with 2 active treatments}.
\bjournal{Biostatistics}
\bvolume{9}
\bpages{277--289}.
\end{barticle}
\endbibitem

%
%
\bibitem[\protect\citeauthoryear{Rubin}{1978a}]{rubinbayesian1978}
\begin{barticle}[author]
\bauthor{\bsnm{Rubin},~\bfnm{Donald~B.}\binits{D.~B.}}
(\byear{1978}a).
\btitle{Bayesian inference for causal effects: The role of randomization}.
\bjournal{Ann. Statist.}
\bvolume{6}
\bpages{34--58}.
\end{barticle}
\MR{0472152}
\endbibitem

%
%
\bibitem[\protect\citeauthoryear{Rubin}{1978b}]{rubinmultiple1978}
\begin{bmisc}[author]
\bauthor{\bsnm{Rubin},~\bfnm{D.~B}\binits{D.~B.}}
(\byear{1978}b).
\bhowpublished{Multiple imputations in sample surveys---a phenomenological
{{B}ayesian} approach to nonresponse. In \textit{Proceedings of the Section on Survey
Research Methods}
20--34.  American Statistical Association, Alexandria, VA}.
\end{bmisc}
\endbibitem

%
%
\bibitem[\protect\citeauthoryear{Shepherd
et~al.}{2006}]{shepherdsensitivity2006}
\begin{barticle}[author]
\bauthor{\bsnm{Shepherd},~\bfnm{Bryan~E.}\binits{B.~E.}},
\bauthor{\bsnm{Gilbert},~\bfnm{Peter~B.}\binits{P.~B.}},
\bauthor{\bsnm{Jemiai},~\bfnm{Yannis}\binits{Y.}} \AND
\bauthor{\bsnm{Rotnitzky},~\bfnm{Andrea}\binits{A.}}
(\byear{2006}).
\btitle{Sensitivity analyses comparing outcomes only existing in a subset
selected post-randomization, conditional on covariates, with
application to
{HIV} vaccine trials}.
\bjournal{Biometrics}
\bvolume{62}
\bpages{332--342}.
\end{barticle}
\MR{2236845}
\endbibitem

%
%
\bibitem[\protect\citeauthoryear{Shetty et~al.}{2007}]{shettymandible2007}
\begin{barticle}[author]
\bauthor{\bsnm{Shetty},~\bfnm{Vivek}\binits{V.}},
\bauthor{\bsnm{Atchison},~\bfnm{Kathryn}\binits{K.}},
\bauthor{\bsnm{{Der-Matirosian}},~\bfnm{Claudia}\binits{C.}},
\bauthor{\bsnm{Wang},~\bfnm{Jianming}\binits{J.}} \AND
\bauthor{\bsnm{Belin},~\bfnm{Thomas~R.}\binits{T.~R.}}
(\byear{2007}).
\btitle{The mandible injury severity score: Development and validity}.
\bjournal{Journal of Oral and Maxillofacial Surgery}
\bvolume{65}
\bpages{663--670}.
\end{barticle}
\endbibitem

%
%
\bibitem[\protect\citeauthoryear{Shetty et~al.}{2008}]{shettybenefits2008}
\begin{barticle}[author]
\bauthor{\bsnm{Shetty},~\bfnm{Vivek}\binits{V.}},
\bauthor{\bsnm{Atchison},~\bfnm{Kathryn}\binits{K.}},
\bauthor{\bsnm{Leathers},~\bfnm{Richard}\binits{R.}},
\bauthor{\bsnm{Black},~\bfnm{Edward}\binits{E.}},
\bauthor{\bsnm{Zigler},~\bfnm{Cory}\binits{C.}} \AND
\bauthor{\bsnm{Belin},~\bfnm{Thomas~R}\binits{T.~R.}}
(\byear{2008}).
\btitle{Do the benefits of rigid internal fixation of mandible fractures
justify the added costs? {{R}esults} from a randomized controlled trial}.
\bjournal{Journal of Oral and Maxillofacial Surgery}
\bvolume{66}
\bpages{2203--2212}.
\end{barticle}
\endbibitem

%
%
\bibitem[\protect\citeauthoryear{Zigler and Belin}{2011}]{ziglersupplement2011}
\begin{bmisc}[author]
\bauthor{\bsnm{Zigler},~\bfnm{Corwin~M.}\binits{C.~M.}} \AND
\bauthor{\bsnm{Belin},~\bfnm{Thomas~R.}\binits{T.~R.}}
(\byear{2011}).
\bhowpublished{Supplement to ``The potential for Bias in principal causal effect
estimation when treatment received depends on a key covariate.'' \href{http://dx.doi.org/10.1214/11-AOAS477SUPP}{DOI:10.1214/11-AOAS477SUPP}}.
\end{bmisc}
\endbibitem

\end{thebibliography}
\end{document}